\documentclass[a4paper]{spie}  
\usepackage[]{graphicx}
\usepackage{color}
\usepackage{amssymb}
\usepackage{amsmath}

\newcommand{\ket}[1]{\vert {#1} \rangle}
\newcommand{\bra}[1]{\langle {#1} \vert}
\newcommand{\dotp}[2]{\langle {#1}\vert{#2} \rangle}
\newcommand{\dyad}[2]{\vert{#1}\rangle\langle{#2}\vert}
\title{
Quasi-Bell entangled coherent states and its quantum discrimination problem in the presence of thermal noise
} 

\author{Kentaro Kato 
\skiplinehalf
Quantum Communication Research Center,\\
Quantum ICT Research Institute,
Tamagawa University, \\
Tamagawa-gakuen 6-1-1, Machida, Tokyo 194-8610 JAPAN
}

\authorinfo{Further author information: E-mail: kkatop@lab.tamagawa.ac.jp}

\begin{document} 
\maketitle 




\begin{abstract}
The so-called quasi-Bell entangled coherent states in a thermal environment are studied. 
In the analysis, we assume thermal noise affects only one of the two modes of each state. First the matrix representation of the density operators of the quasi-Bell entangled coherent states in a thermal environment is derived. Secondly we investigate the entanglement property of one of the quasi-Bell entangled coherent states with thermal noise. At that time a lower bound of the entanglement of formation for the state is computed. Thirdly the minimax discrimination problem for two cases of the binary set of the quasi-Bell entangled coherent states with thermal noise is considered, and the error probabilities of the minimax discrimination for the two cases are computed with the help of Helstrom's algorithm for finding the Bayes optimal error probability of binary states.
\end{abstract}
\section{Introduction}
The choice or design of quantum states for information-processing resource is one of important issues to build a useful quantum information system that breaks the performance limit of classical schemes in information-communication technologies or that possesses quite new functions. Entanglement may be one of potential phenomena and indeed various types of entangled states have been widely discussed from the point of view of both theory and experiment and from fundamental properties to its applications. Toward developing macroscopic quantum information-communication technologies, we are interested in entangled coherent states \cite{San12,San13}. 
Above all, our interest is focused on the so-called quasi-Bell entangled coherent states in this paper.
The quasi-Bell entangled coherent states are defined as a set of the following four superpositions of two-mode coherent states:
$\ket{\alpha,\pm\alpha}_{12}+\ket{-\alpha,\mp\alpha}_{12}$
and
$\ket{\alpha,\pm\alpha}_{12}-\ket{-\alpha,\mp\alpha}_{12}$ \cite{HiS01}.

In an ideal situation, it has been clarified that the quasi-Bell entangled coherent states are applicable to several applications. For example, teleportation and computation by the quasi-Bell entangled coherent states were already proposed \cite{EnH01,Wan01,JKL01,JeK02}. The error-free quantum reading scheme was proposed by Hirota \cite{Hir11}, where phase-shift keying (PSK) format was used for its signal modulation.  In order to realise such information-communication systems, more precise analysis of the states in various realistic situations is needed. In other words, our central concern is whether or not the quasi-Bell entangled coherent states are still applicable to information-communication technologies even in practical situations. For this question, the effect of channel loss to the quasi-Bell entangled coherent states was reported by present author in the context of the quantum reading in SPIE Optics+Photonics 2013 \cite{KaH13}. In line with this thought, we aim to clarify the effect of thermal noise for the quasi-Bell entangled coherent states.

This paper is organized as follows. Section 2 reviews the basics of the quasi-Bell entangled coherent states when there is no noise. In Section 3, we derive the matrix representation of the density operators of the quasi-Bell entangled coherent states in a thermal environment by using the thermalizing operator. The degree of entanglement of one of the quasi-Bell entangled coherent states with thermal noise is discussed in Section 4, and an optimal quantum discrimination problem for two cases of the binary set of the quasi-Bell entangled coherent states with thermal noise is considered in Section 5. In these sections, we execute computer simulations by using the matrix representation mentioned in Section 3.
Section 6 summarizes our results obtained in this paper.

\section{Quasi-Bell Entangled Coherent States without thermal noise}
Consider two modes of electromagnetic field with corresponding annihilation operators $\hat{a}_1$ and $\hat{a}_2$.
Two-mode coherent states are defined by 
$
	\ket{\alpha_1,\alpha_2}_{12}
	=\hat{D}_1(\alpha_1)\hat{D}_2(\alpha_2)\ket{0,0}_{12},
$
where $\ket{0,0}_{12}$ is the two-mode vacuum state and 
$\hat{D}_i(\alpha_i) = \exp[\alpha_i\hat{a}_i^\dagger-\alpha^\ast \hat{a}_i]$ is the displacement operator of the mode $i$ ($i=1,2$). 
The state $\ket{\alpha_1,\alpha_2}_{12}$ can be expressed into the form
\begin{equation}
	\ket{\alpha_1,\alpha_2}_{12}
	=
	\exp[-\frac{|\alpha_1|^2}{2}-\frac{|\alpha_2|^2}{2}]
	\sum_{n_1=0}^\infty
	\sum_{n_2=0}^\infty
	\frac{\alpha_1^{n_1}\alpha_2^{n_2}}{\sqrt{n_1!n_2!}}
	\ket{n_1,n_2}_{12},
\end{equation}
where $\ket{n_1,n_2}_{12}$ are two-mode Fock states.
The quasi-Bell entangled coherent states are defined by the following superpositions of two-mode coherent states \cite{HiS01}.
\begin{eqnarray}
	\ket{\boldsymbol{\Phi}_1}_{\rm 12}
	&=&
	\mathcal{N}_+
	\left(
	\ket{ \beta, \beta}_{\rm 12}
	+
	\ket{-\beta,-\beta}_{\rm 12}
	\right),
	\\
	\ket{\boldsymbol{\Phi}_2}_{\rm 12}
	&=&
	\mathcal{N}_-
	\left(
	\ket{ \beta, \beta}_{\rm 12}
	-
	\ket{-\beta,-\beta}_{\rm 12}
	\right),
	\\
	\ket{\boldsymbol{\Phi}_3}_{\rm 12}
	&=&
	\mathcal{N}_+
	\left(
	\ket{ \beta,-\beta}_{\rm 12}
	+
	\ket{-\beta, \beta}_{\rm 12}
	\right),
	\\
	\ket{\boldsymbol{\Phi}_4}_{\rm 12}
	&=&
	\mathcal{N}_-
	\left(
	\ket{ \beta,-\beta}_{\rm 12}
	-
	\ket{-\beta, \beta}_{\rm 12}
	\right),
\end{eqnarray}
where 
$\mathcal{N}_{\pm}={1}/{\sqrt{2(1\pm\exp[-4|\beta|^2])}}$
is the normalizing factors. 
From these expressions of the states, one can find the relations
\begin{equation}
	\ket{\boldsymbol{\Phi}_3}_{\rm 12}
	=
	(\hat{1}_1\otimes\hat{R}_2(\pi))
	\ket{\boldsymbol{\Phi}_1}_{\rm 12},
	\qquad
	\ket{\boldsymbol{\Phi}_4}_{\rm 12}
	=
	(\hat{1}_1\otimes\hat{R}_2(\pi))
	\ket{\boldsymbol{\Phi}_2}_{\rm 12},
	\label{purecasesymmetryrelations}
\end{equation}
where $\hat{R}_2(\pi)=\exp[-{\rm i}\pi\hat{a}_2^\dagger\hat{a}_2]$ is the $\pi$-rotation operator. In the context of optical communications, this property is understood as the PSK format of laser light.

The average numbers of photons for the states $\boldsymbol{\Phi}_1$ and $\boldsymbol{\Phi}_3$ are given by
\begin{eqnarray}
	\langle n(\boldsymbol{\Phi}_1) \rangle
	=
	\langle n(\boldsymbol{\Phi}_3) \rangle
	&=&
	2|\beta|^2\tanh[2|\beta|^2].
\end{eqnarray}
It is easy to see that
$\langle n(\boldsymbol{\Phi}_1) \rangle
=
\langle n(\boldsymbol{\Phi}_3) \rangle \to 0$ 
when $\beta\to 0$.
Similarly, the average numbers of photons for the states $\boldsymbol{\Phi}_2$ and $\boldsymbol{\Phi}_4$ are given by
\begin{eqnarray}
	\langle n(\boldsymbol{\Phi}_2) \rangle
	=
	\langle n(\boldsymbol{\Phi}_4) \rangle
	&=&
	2|\beta|^2\coth[2|\beta|^2],
\end{eqnarray}
and
$\langle n(\boldsymbol{\Phi}_2) \rangle
	=
	\langle n(\boldsymbol{\Phi}_4) \rangle \to 1$ when $\beta\to 0$.
Symbol $\langle n(\boldsymbol{\Phi}_i) \rangle$ is abbreviated to $\langle n \rangle$ if it is clear from the context.

The set of the inner products of the quasi-Bell entangled coherent states forms the following Gram matrix.
\begin{equation}
	{\sf G}
	=
	\left[
	\begin{array}{cccc}
	{}_{\rm 12}\dotp{\boldsymbol{\Phi}_1}{\boldsymbol{\Phi}_1}_{\rm 12} &
	{}_{\rm 12}\dotp{\boldsymbol{\Phi}_1}{\boldsymbol{\Phi}_2}_{\rm 12} &
	{}_{\rm 12}\dotp{\boldsymbol{\Phi}_1}{\boldsymbol{\Phi}_3}_{\rm 12} &
	{}_{\rm 12}\dotp{\boldsymbol{\Phi}_1}{\boldsymbol{\Phi}_4}_{\rm 12} \\
	{}_{\rm 12}\dotp{\boldsymbol{\Phi}_2}{\boldsymbol{\Phi}_1}_{\rm 12} &
	{}_{\rm 12}\dotp{\boldsymbol{\Phi}_2}{\boldsymbol{\Phi}_2}_{\rm 12} &
	{}_{\rm 12}\dotp{\boldsymbol{\Phi}_2}{\boldsymbol{\Phi}_3}_{\rm 12} &
	{}_{\rm 12}\dotp{\boldsymbol{\Phi}_2}{\boldsymbol{\Phi}_4}_{\rm 12} \\
	{}_{\rm 12}\dotp{\boldsymbol{\Phi}_3}{\boldsymbol{\Phi}_1}_{\rm 12} &
	{}_{\rm 12}\dotp{\boldsymbol{\Phi}_3}{\boldsymbol{\Phi}_2}_{\rm 12} &
	{}_{\rm 12}\dotp{\boldsymbol{\Phi}_3}{\boldsymbol{\Phi}_3}_{\rm 12} &
	{}_{\rm 12}\dotp{\boldsymbol{\Phi}_3}{\boldsymbol{\Phi}_4}_{\rm 12} \\
	{}_{\rm 12}\dotp{\boldsymbol{\Phi}_4}{\boldsymbol{\Phi}_1}_{\rm 12} &
	{}_{\rm 12}\dotp{\boldsymbol{\Phi}_4}{\boldsymbol{\Phi}_2}_{\rm 12} &
	{}_{\rm 12}\dotp{\boldsymbol{\Phi}_4}{\boldsymbol{\Phi}_3}_{\rm 12} &
	{}_{\rm 12}\dotp{\boldsymbol{\Phi}_4}{\boldsymbol{\Phi}_4}_{\rm 12} 
	\end{array}
	\right]
	=
	\left[
	\begin{array}{cccc}
	1 & 0 & K_{13} & 0\\
	0 & 1 & 0 & 0\\
	K_{13} & 0 & 1 & 0\\
	0 & 0 & 0 & 1
	\end{array}
	\right],
	\label{grammatrixpure}
\end{equation}
where the entry $K_{13}={\rm sech}[2|\beta|^2]$. 
From this matrix, it is noticed that the states 
$\boldsymbol{\Phi}_1$ and 
$\boldsymbol{\Phi}_3$ are non-orthogonal, 
while any other pair is orthogonal. The orthogonality of the states $\boldsymbol{\Phi}_2$ and $\boldsymbol{\Phi}_4$ is utilized to propose an error-free binary PSK quantum reading \cite{Hir11}. Indeed, the minimum probability of error (in terms of the minimax discrimination problem \cite{HiI82,Kat12a,NKU15}) for the states $\boldsymbol{\Phi}_2$ and $\boldsymbol{\Phi}_4$ is 
\begin{eqnarray}
	\bar{P}_{\rm e}^{\circ}
	(\boldsymbol{\Phi}_2,\boldsymbol{\Phi}_4)
	&=&
	\max_{\{p_2+p_4=1;p_2\geq 0;p_4\geq 0\}}
	\left[
	\frac{1}{2}
	\left(
	1-\sqrt{1-4p_2p_4
	\bigl|
	{}_{\rm 12}\dotp{\boldsymbol{\Phi}_2}{\boldsymbol{\Phi}_4}_{\rm 12}
	\bigr|^2}
	\right)
	\right]
	=0,
\end{eqnarray}
while the minimum probability of error for the states $\boldsymbol{\Phi}_1$ and $\boldsymbol{\Phi}_3$ is
\begin{eqnarray}
	\bar{P}_{\rm e}^{\circ}
	(\boldsymbol{\Phi}_1,\boldsymbol{\Phi}_3)
	&=&
	\max_{\{p_1+p_3=1;p_1\geq 0;p_3\geq 0\}}
	\left[
	\frac{1}{2}
	\left(
	1-\sqrt{1-4p_1p_3
	\bigl|
	{}_{\rm 12}\dotp{\boldsymbol{\Phi}_1}{\boldsymbol{\Phi}_3}_{\rm 12}
	\bigr|^2}
	\right)
	\right]
	=
	\frac{\exp[-4|\beta|^2]}{1+\exp[-4|\beta|^2]}.
\end{eqnarray}

It is well known that the states $\boldsymbol{\Phi}_2$ and $\boldsymbol{\Phi}_4$ are maximally entangled states \cite{HiS01}. 
The degrees of entanglement of the quasi-Bell entangled coherent states are given 
by
\begin{eqnarray}
	{E}(\ket{\boldsymbol{\Phi}_1}_{\rm 12})
	=
	{E}(\ket{\boldsymbol{\Phi}_3}_{\rm 12})
	&=&
	-\left(\frac{1+K_{13}}{2}\right)
	\log_2
	\left[
	\frac{1+K_{13}}{2}
	\right]
	-\left(\frac{1-K_{13}}{2}\right)
	\log_2
	\left[
	\frac{1-K_{13}}{2}
	\right],
	\label{entropyofentanglement13}
	\\
	{E}(\ket{\boldsymbol{\Phi}_2}_{\rm 12})
	=
	{E}(\ket{\boldsymbol{\Phi}_4}_{\rm 12})
	&=&
	1,
	\label{maximallyentanglement}
\end{eqnarray}
where we have used the {entropy of entanglement} \cite{BaP89,BBPS96} defined by
\begin{equation}
	E(\ket{\boldsymbol{\Phi}}_{\rm 12})
	=
	-{\rm Tr}_{\rm 1}
	\left(
	\hat{{\rho}}_{\rm 1}
	\log_2
	\hat{{\rho}}_{\rm 1}
	\right)
	=
	-{\rm Tr}_{\rm 2}
	\left(
	\hat{{\rho}}_{\rm 2}
	\log_2
	\hat{{\rho}}_{\rm 2}
	\right)
\end{equation}
with the reduced density operators 
$
	\hat{{\rho}}_{\rm 1}
	=
	{\rm Tr}_{\rm 2}
	\ket{\boldsymbol{\Phi}}_{\rm 12}
	\bra{\boldsymbol{\Phi}}
$ and $
	\hat{{\rho}}_{\rm 2}
	=
	{\rm Tr}_{\rm 1}
	\ket{\boldsymbol{\Phi}}_{\rm 12}
	\bra{\boldsymbol{\Phi}}
$.
Eq.(\ref{maximallyentanglement}) shows that $\boldsymbol{\Phi}_2$ and $\boldsymbol{\Phi}_4$ are maximally entangled when there is no noise. 

\section{Quasi-Bell entangled coherent states in a thermal environment}
In this section, we give the matrix representation of the density operators of the quasi-Bell entangled coherent states in a thermal environment. For this aim, we use the
thermalizing operator \cite{TaU75, BaK85} $\hat{\bf T}(\theta)$ defined by
\begin{equation}
	\hat{\bf T}(\theta)
	=
	\exp[
	-\theta(\hat{a}\tilde{a}-\hat{a}^\dagger\tilde{a}^\dagger)
	].
	\label{original:thermalizingop}
\end{equation}
In this definition, 
$\hat{a}$ is the annihilation operator of the real mode to be primary investigated,
and $\tilde{a}$ is the annihilation operator of the fictitious mode that satisfy
$[\hat{a},\tilde{a}]=0=[\hat{a},\tilde{a}^\dagger]$
and
$[\tilde{a},\tilde{a}^\dagger]=\hat{1}$. 
Further, 
$
	\cosh\theta={1}/{\sqrt{1-{\rm e}^{-\omega/k_{\rm B}T}}},
$ 
and
$
	\sinh\theta={1}/{\sqrt{{\rm e}^{\omega/k_{\rm B}T}-1}}
$
with 
frequency $\omega$, Boltzmann constant $k_{\rm B}$, and temperature $T$. 
By using the thermalizing operator, various types of thermal states can be considered: 
(i) thermal vacuum state \cite{TaU75,BaK85},
$
	\ket{\theta;0,\tilde{0}}=\hat{\bf T}(\theta)\ket{0,\tilde{0}}
$, where $\ket{0,\tilde{0}}$ is the zero-temperature vacuum, 
(ii) thermal coherent state \cite{BaK85,FeC88,MaR89,OVMR91},
$
	\ket{\theta;\alpha_1,\alpha_2}
	=
	\hat{\bf T}(\theta)\ket{\alpha_1,\alpha_2}
	=
	\hat{\bf T}(\theta)\hat{D}(\alpha_1)\tilde{D}(\alpha_2)\ket{0,\tilde{0}}
$
(iii) coherent thermal state \cite{FeC88},
$
	\ket{\alpha_1^\prime,\alpha_2^\prime;\theta}
	=
	\hat{D}(\alpha_1^\prime)\tilde{D}(\alpha_2^\prime)\hat{\bf T}(\theta)\ket{0,\tilde{0}}
$
(iv) thermal squeezed state --- thermalizing a squeezed state --- \cite{FeC88,KMRU89,MaR89,Lee90,OVMR91},
(v) squeezed thermal state --- squeezing a thermal state --- \cite{FeC88,Lee90}, 
and so on \cite{Lu99,Lu00}. Thus the thermalizing operator provides a useful tool in the analysis of optical quantum states in a thermal environment.

Here let us describe the model to be investigated in this study. 
We assume thermal noise affects only mode 2. Then the quasi-Bell entangled coherent states in a thermal environment is formally written as follows.
\begin{eqnarray}
	\hat{\boldsymbol{\rho}}_{12}(\boldsymbol{\Phi}_1)
	&=&
	{\rm Tr}_3
	\Bigl[
	\Bigl\{
	\Bigl(
	\hat{1}_1\otimes\hat{\bf T}_{23}(\theta)
	\Bigr)
	\ket{\boldsymbol{\Phi}_1}_{12}
	\ket{\tilde{0}}_3
	\Bigr\}
	\Bigl\{
	{}_{12}\bra{\boldsymbol{\Phi}_1}
	{}_3\bra{\tilde{0}}
	\Bigl(
	\hat{1}_1\otimes
	\hat{\bf T}_{23}^\dagger(\theta)
	\Bigr)
	\Bigr\}
	\Bigr],
\end{eqnarray}
\begin{eqnarray}
	\hat{\boldsymbol{\rho}}_{12}(\boldsymbol{\Phi}_2)
	&=&
	{\rm Tr}_3
	\Bigl[
	\Bigl\{
	\Bigl(
	\hat{1}_1\otimes\hat{\bf T}_{23}(\theta)
	\Bigr)
	\ket{\boldsymbol{\Phi}_2}_{12}
	\ket{\tilde{0}}_3
	\Bigr\}
	\Bigl\{
	{}_{12}\bra{\boldsymbol{\Phi}_2}
	{}_3\bra{\tilde{0}}
	\Bigl(
	\hat{1}_1\otimes
	\hat{\bf T}_{23}^\dagger(\theta)
	\Bigr)
	\Bigr\}
	\Bigr],
\end{eqnarray}
\begin{eqnarray}
	\hat{\boldsymbol{\rho}}_{12}(\boldsymbol{\Phi}_3)
	&=&
	{\rm Tr}_3
	\Bigl[
	\Bigl\{
	\Bigl(
	\hat{1}_1\otimes\hat{\bf T}_{23}(\theta)
	\Bigr)
	\ket{\boldsymbol{\Phi}_3}_{12}
	\ket{\tilde{0}}_3
	\Bigr\}
	\Bigl\{
	{}_{12}\bra{\boldsymbol{\Phi}_3}
	{}_3\bra{\tilde{0}}
	\Bigl(
	\hat{1}_1\otimes
	\hat{\bf T}_{23}^\dagger(\theta)
	\Bigr)
	\Bigr\}
	\Bigr],
\end{eqnarray}
\begin{eqnarray}
	\hat{\boldsymbol{\rho}}_{12}(\boldsymbol{\Phi}_4)
	&=&
	{\rm Tr}_3
	\Bigl[
	\Bigl\{
	\Bigl(
	\hat{1}_1\otimes\hat{\bf T}_{23}(\theta)
	\Bigr)
	\ket{\boldsymbol{\Phi}_4}_{12}
	\ket{\tilde{0}}_3
	\Bigr\}
	\Bigl\{
	{}_{12}\bra{\boldsymbol{\Phi}_4}
	{}_3\bra{\tilde{0}}
	\Bigl(
	\hat{1}_1\otimes
	\hat{\bf T}_{23}^\dagger(\theta)
	\Bigr)
	\Bigr\}
	\Bigr],
\end{eqnarray}
where mode 3 is the fictitious mode for calculating the effect of thermal noise that affects only on mode 2, so that we have set
$
\hat{\bf T}_{23}(\theta)
=
\exp
[
-\theta(\hat{a}_2\tilde{a}_3-\hat{a}_2^\dagger\tilde{a}_3^\dagger)
]
$.
By using the overcompleteness of two-mode coherent states 
$
	({1}/{\pi^2})\int 
	{\rm d}^2\gamma_1
	{\rm d}^2\gamma_2
	\dyad{\gamma_1,\gamma_2}{\gamma_1,\gamma_2}
	=
	\hat{\bf 1}_{12}
$,
the four states mentioned above are rewritten as follows.
\begin{eqnarray}
\hat{\boldsymbol{\rho}}_{12}(\boldsymbol{\Phi}_1)
&=&
\frac{\mathcal{N}_+^2}{\pi^4}
\int\!\!\!
\int\!\!\!
\int\!\!\!
\int
{\rm d}^2\gamma_1
{\rm d}^2\gamma_1^\prime
{\rm d}^2\gamma_2
{\rm d}^2\gamma_2^\prime
[
F_1(\beta,\beta)F_2(\beta,\beta)
+
F_1(\beta,-\beta)F_2(\beta,-\beta)
\nonumber\\
&&
\qquad\qquad\qquad\qquad
+
F_1(-\beta,\beta)F_2(-\beta,\beta)
+
F_1(-\beta,-\beta)F_2(-\beta,-\beta)
]
\ket{\gamma_1,\gamma_2}_{12}\bra{\gamma_1^\prime,\gamma_2^\prime},
\label{sekibunkei01}
\end{eqnarray}
\begin{eqnarray}
\hat{\boldsymbol{\rho}}_{12}(\boldsymbol{\Phi}_2)
&=&
\frac{\mathcal{N}_-^2}{\pi^4}
\int\!\!\!
\int\!\!\!
\int\!\!\!
\int
{\rm d}^2\gamma_1
{\rm d}^2\gamma_1^\prime
{\rm d}^2\gamma_2
{\rm d}^2\gamma_2^\prime
\Bigl[
F_1(\beta,\beta)F_2(\beta,\beta)
-
F_1(\beta,-\beta)F_2(\beta,-\beta)
\nonumber\\
&&
\qquad\qquad\qquad\qquad
-
F_1(-\beta,\beta)F_2(-\beta,\beta)
+
F_1(-\beta,-\beta)F_2(-\beta,-\beta)
\Bigr]
\ket{\gamma_1,\gamma_2}_{12}\bra{\gamma_1^\prime,\gamma_2^\prime},
\label{sekibunkei02}
\end{eqnarray}
\begin{eqnarray}
\hat{\boldsymbol{\rho}}_{12}(\boldsymbol{\Phi}_3)
&=&
\frac{\mathcal{N}_+^2}{\pi^4}
\int\!\!\!
\int\!\!\!
\int\!\!\!
\int
{\rm d}^2\gamma_1
{\rm d}^2\gamma_1^\prime
{\rm d}^2\gamma_2
{\rm d}^2\gamma_2^\prime
\Bigl[
F_1(\beta,\beta)F_2(-\beta,-\beta)
+
F_1(\beta,-\beta)F_2(-\beta,\beta)
\nonumber\\
&&
\qquad\qquad\qquad\qquad
+
F_1(-\beta,\beta)F_2(\beta,-\beta)
+
F_1(-\beta,-\beta)F_2(\beta,\beta)
\Bigr]
\ket{\gamma_1,\gamma_2}_{12}\bra{\gamma_1^\prime,\gamma_2^\prime},
\label{sekibunkei03}
\end{eqnarray}
\begin{eqnarray}
\hat{\boldsymbol{\rho}}_{12}(\boldsymbol{\Phi}_4)
&=&
\frac{\mathcal{N}_-^2}{\pi^4}
\int\!\!\!
\int\!\!\!
\int\!\!\!
\int
{\rm d}^2\gamma_1
{\rm d}^2\gamma_1^\prime
{\rm d}^2\gamma_2
{\rm d}^2\gamma_2^\prime
\Bigl[
F_1(\beta,\beta)F_2(-\beta,-\beta)
-
F_1(\beta,-\beta)F_2(-\beta,\beta)
\nonumber\\
&&
\qquad\qquad\qquad\qquad
-
F_1(-\beta,\beta)F_2(\beta,-\beta)
+
F_1(-\beta,-\beta)F_2(\beta,\beta)
\Bigr]
\ket{\gamma_1,\gamma_2}_{12}\bra{\gamma_1^\prime,\gamma_2^\prime},
\label{sekibunkei04}
\end{eqnarray}
where
\begin{eqnarray}
	F_1(\beta_L,\beta_R)
	&=&
	\exp\left[
	-|\beta|^2
	-\frac{1}{2}|\gamma_1|^2+\beta_L \gamma_1^\ast
	-\frac{1}{2}|\gamma_1^\prime|^2+\beta^\ast_R \gamma_1^\prime
	\right],
\\
F_2(\beta_L,\beta_R)
&=&
\frac{1}{\cosh^2\theta}
\exp\left[-|\beta|^2
-\frac{1}{2}|\gamma_2|^2+\frac{\beta_L}{\cosh\theta}\gamma_2^\ast
-\frac{1}{2}|\gamma_2^\prime|^2+\frac{\beta_R^\ast}{\cosh\theta}\gamma_2^{\prime}
+\gamma_2^\ast\gamma_2^\prime\tanh^2\theta
\right],
\end{eqnarray}
for $\beta_L,\beta_R\in\{-\beta,\beta\}$. 
From the expressions of Eqs.(\ref{sekibunkei01})-(\ref{sekibunkei04}), one can find  symmetric relations of the states. For $\hat{\boldsymbol{\rho}}_{12}(\boldsymbol{\Phi}_1)$ and $\hat{\boldsymbol{\rho}}_{12}(\boldsymbol{\Phi}_3)$, we have 
$
	\hat{\boldsymbol{\rho}}_{12}(\boldsymbol{\Phi}_1)
	=
	(\hat{1}_1\otimes \hat{R}_2(\pi))
	\hat{\boldsymbol{\rho}}_{12}(\boldsymbol{\Phi}_3)
	(\hat{1}_1\otimes \hat{R}_2(\pi))^\dagger
$. 
Similarly, 
$
	\hat{\boldsymbol{\rho}}_{12}(\boldsymbol{\Phi}_2)
	=
	(\hat{1}_1\otimes \hat{R}_2(\pi))
	\hat{\boldsymbol{\rho}}_{12}(\boldsymbol{\Phi}_4)
	(\hat{1}_1\otimes \hat{R}_2(\pi))^\dagger
$
for $\hat{\boldsymbol{\rho}}_{12}(\boldsymbol{\Phi}_2)$ and $\hat{\boldsymbol{\rho}}_{12}(\boldsymbol{\Phi}_4)$. As we will see later, this type of symmetry of the states reduces an optimization step in the minimax discrimination problem.

To execute computer simulations, we rewrite the density operators of Eqs.(\ref{sekibunkei01})-(\ref{sekibunkei04}) into the corresponding matrix representation by Fock states $\{\ket{n_1,n_2}_{12}\}$. The matrix representation of $\hat{\boldsymbol{\rho}}_{12}(\boldsymbol{\Phi}_i)$ in the basis $\{\ket{n_1,n_2}_{12}\}$ is given as follows. 
\begin{eqnarray}
\hat{\boldsymbol{\rho}}_{12}(\boldsymbol{\Phi}_1)
&=&
\mathcal{N}_+^2
\sum_{m_1=0}^\infty
\sum_{m_2=0}^\infty
\sum_{n_1=0}^\infty
\sum_{n_2=0}^\infty
\sum_{k=0}^\infty
\Bigl[
g_1(\beta,m_1)
g_1^\ast(\beta,n_1)
g_{23}(\beta,m_2,k)
g_{23}^\ast(\beta,n_2,k)
\nonumber\\
&&
\qquad\qquad\qquad\qquad\qquad\qquad
+
g_1(\beta,m_1)
g_1^\ast(-\beta,n_1)
g_{23}(\beta,m_2,k)
g_{23}^\ast(-\beta,n_2,k)
\nonumber\\
&&
\qquad\qquad\qquad\qquad\qquad\qquad
+
g_1(-\beta,m_1)
g_1^\ast(\beta,n_1)
g_{23}(-\beta,m_2,k)
g_{23}^\ast(\beta,n_2,k)
\nonumber\\
&&
\qquad\qquad\qquad\qquad\qquad\qquad
+
g_1(-\beta,m_1)
g_1^\ast(-\beta,n_1)
g_{23}(-\beta,m_2,k)
g_{23}^\ast(-\beta,n_2,k)
\Bigr]
\nonumber\\
&&
\qquad\qquad\qquad\qquad\qquad\qquad
\times
\ket{m_1,m_2}_{12}\bra{n_1,n_2},
\label{matrixrepresentation001}
\end{eqnarray}
\begin{eqnarray}
\hat{\boldsymbol{\rho}}_{12}(\boldsymbol{\Phi}_2)
&=&
\mathcal{N}_-^2
\sum_{m_1=0}^\infty
\sum_{m_2=0}^\infty
\sum_{n_1=0}^\infty
\sum_{n_2=0}^\infty
\sum_{k=0}^\infty
\Bigl[
g_1(\beta,m_1)
g_1^\ast(\beta,n_1)
g_{23}(\beta,m_2,k)
g_{23}^\ast(\beta,n_2,k)
\nonumber\\
&&
\qquad\qquad\qquad\qquad\qquad\qquad
-
g_1(\beta,m_1)
g_1^\ast(-\beta,n_1)
g_{23}(\beta,m_2,k)
g_{23}^\ast(-\beta,n_2,k)
\nonumber\\
&&
\qquad\qquad\qquad\qquad\qquad\qquad
-
g_1(-\beta,m_1)
g_1^\ast(\beta,n_1)
g_{23}(-\beta,m_2,k)
g_{23}^\ast(\beta,n_2,k)
\nonumber\\
&&
\qquad\qquad\qquad\qquad\qquad\qquad
+
g_1(-\beta,m_1)
g_1^\ast(-\beta,n_1)
g_{23}(-\beta,m_2,k)
g_{23}^\ast(-\beta,n_2,k)
\Bigr]
\nonumber\\
&&
\qquad\qquad\qquad\qquad\qquad\qquad
\times
\ket{m_1,m_2}_{12}\bra{n_1,n_2},
\label{matrixrepresentation002}
\end{eqnarray}
\begin{eqnarray}
\hat{\boldsymbol{\rho}}_{12}(\boldsymbol{\Phi}_3)
&=&
\mathcal{N}_+^2
\sum_{m_1=0}^\infty
\sum_{m_2=0}^\infty
\sum_{n_1=0}^\infty
\sum_{n_2=0}^\infty
\sum_{k=0}^\infty
\Bigl[
g_1(\beta,m_1)
g_1^\ast(\beta,n_1)
g_{23}(-\beta,m_2,k)
g_{23}^\ast(-\beta,n_2,k)
\nonumber\\
&&
\qquad\qquad\qquad\qquad\qquad\qquad
+
g_1(\beta,m_1)
g_1^\ast(-\beta,n_1)
g_{23}(-\beta,m_2,k)
g_{23}^\ast(\beta,n_2,k)
\nonumber\\
&&
\qquad\qquad\qquad\qquad\qquad\qquad
+
g_1(-\beta,m_1)
g_1^\ast(\beta,n_1)
g_{23}(\beta,m_2,k)
g_{23}^\ast(-\beta,n_2,k)
\nonumber\\
&&
\qquad\qquad\qquad\qquad\qquad\qquad
+
g_1(-\beta,m_1)
g_1^\ast(-\beta,n_1)
g_{23}(\beta,m_2,k)
g_{23}^\ast(\beta,n_2,k)
\Bigr]
\nonumber\\
&&
\qquad\qquad\qquad\qquad\qquad\qquad
\times
\ket{m_1,m_2}_{12}\bra{n_1,n_2},
\label{matrixrepresentation003}
\end{eqnarray}
\begin{eqnarray}
\hat{\boldsymbol{\rho}}_{12}(\boldsymbol{\Phi}_4)
&=&
\mathcal{N}_-^2
\sum_{m_1=0}^\infty
\sum_{m_2=0}^\infty
\sum_{n_1=0}^\infty
\sum_{n_2=0}^\infty
\sum_{k=0}^\infty
\Bigl[
g_1(\beta,m_1)
g_1^\ast(\beta,n_1)
g_{23}(-\beta,m_2,k)
g_{23}^\ast(-\beta,n_2,k)
\nonumber\\
&&
\qquad\qquad\qquad\qquad\qquad\qquad
-
g_1(\beta,m_1)
g_1^\ast(-\beta,n_1)
g_{23}(-\beta,m_2,k)
g_{23}^\ast(\beta,n_2,k)
\nonumber\\
&&
\qquad\qquad\qquad\qquad\qquad\qquad
-
g_1(-\beta,m_1)
g_1^\ast(\beta,n_1)
g_{23}(\beta,m_2,k)
g_{23}^\ast(-\beta,n_2,k)
\nonumber\\
&&
\qquad\qquad\qquad\qquad\qquad\qquad
+
g_1(-\beta,m_1)
g_1^\ast(-\beta,n_1)
g_{23}(\beta,m_2,k)
g_{23}^\ast(\beta,n_2,k)
\Bigr]
\nonumber\\
&&
\qquad\qquad\qquad\qquad\qquad\qquad
\times
\ket{m_1,m_2}_{12}\bra{n_1,n_2},
\label{matrixrepresentation004}
\end{eqnarray}
\begin{figure}[b!]
\vspace{2em}
\hrule
\vspace{2em}
\centerline{
	\includegraphics[scale = 1]{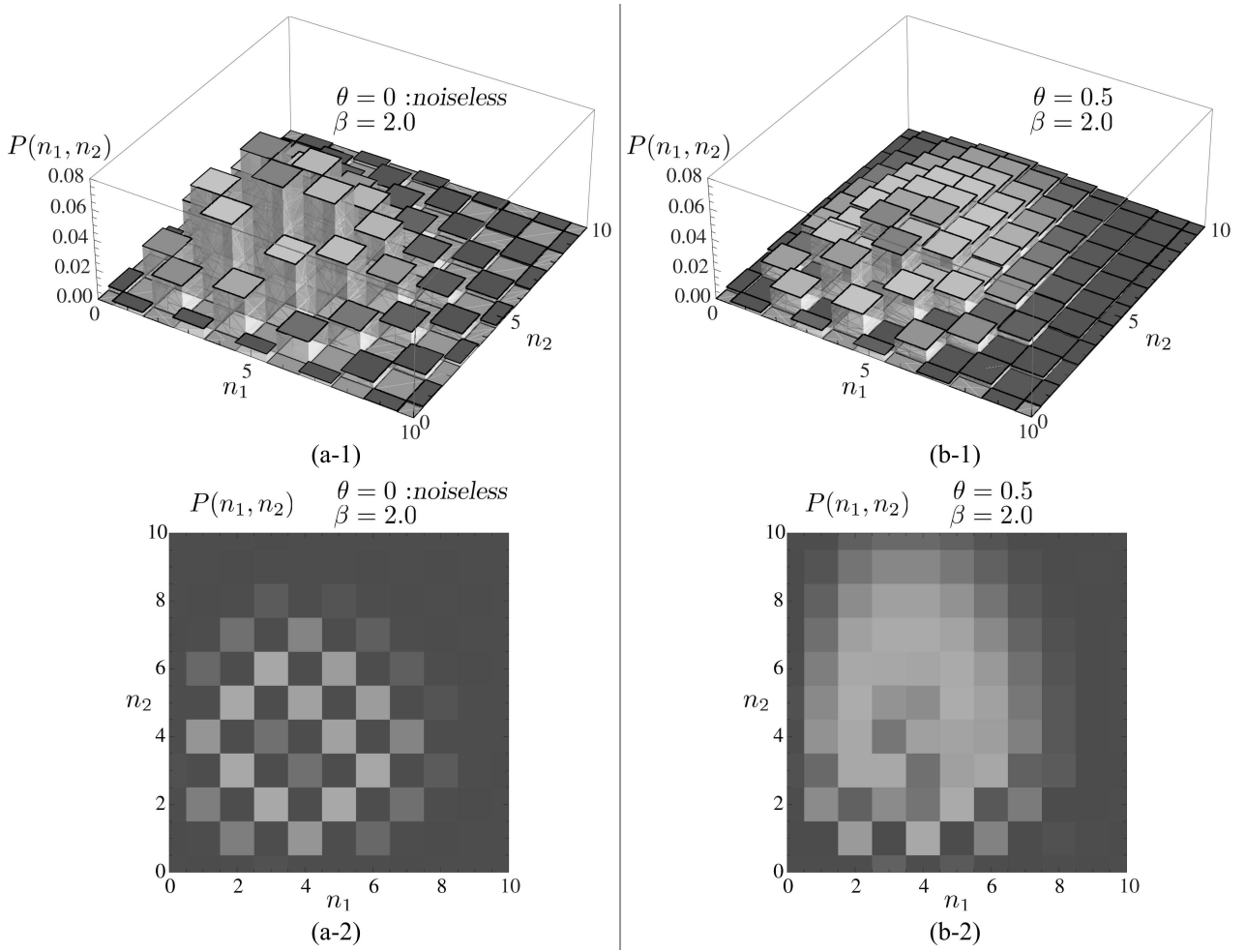}
}
\caption{
Photon-number distribution $P(n_1,n_2)$ of the state $\boldsymbol{\Phi}_2$. 
(a-1) $P(n_1,n_2)$ for $\theta=0$ and $\beta=2.0$. 
(a-2) Contour plot of (a-1). 
(b-1) $P(n_1,n_2)$ for $\theta=0.5$ and $\beta=2.0$. 
(b-2) Contour plot of (b-1). 
}
\label{fig1}
\end{figure}
where 
\begin{eqnarray}
	g_1(\beta^\prime,k_1)
	&=&
	{}_{1}\dotp{k_1}{\beta^\prime}_1
	=
	\exp[-\frac{|\beta^\prime|^2}{2}]
	\frac{\beta^{\prime k_1}}{\sqrt{{k_1}!}},
	\\
	g_{23}(\beta^\prime,k_2,k_3)
	&=&
	{}_{23}
	\bra{k_2,k_3}
	\hat{\bf T}_{23}(\theta)
	\ket{\beta^\prime,\tilde{0}}_{23}
	\nonumber\\
	&=&
	\left\{
	\begin{array}{rl}
	\displaystyle
	\frac{1}{\cosh\theta}
	\exp[-\frac{1}{2}|\beta^\prime|^2]
	\sqrt{\frac{k_3!}{k_2!}}
	{k_2 \choose k_3}
	\beta^{\prime k_2-k_3}
	\frac{(\sinh\theta)^{k_3}}{(\cosh\theta)^{k_2}},
	& \mbox{if } k_2\geq k_3\\
	& \\
	0, & \mbox{if } k_2< k_3
	\end{array}
	\right.
	\label{innerproductofthermalandcoherent18}
\end{eqnarray}
and we have applied the calculation result of Ref. 24
to obtain Eq.(\ref{innerproductofthermalandcoherent18}). 
From Eqs.(\ref{matrixrepresentation001})-(\ref{matrixrepresentation004}), the photon-number distributions 
$
	P(n_1,n_2)
	=
	\bra{n_1,n_2}
	\hat{\boldsymbol{\rho}}_{12}(\boldsymbol{\Phi}_i)
	\ket{n_1,n_2}
$ can be derived.
Figure \ref{fig1} shows examples of photon-number distribution $P(n_1,n_2)$ of the state $\boldsymbol{\Phi}_2$ for $\beta=2.0$. 
The graphs (a-1) and (a-2) show the case of $\theta=0$, that is, noiseless case, and the graphs (b-1) and (b-2) show the case of $\theta=0.5$.
When there is no thermal noise ($\theta=0$), the distribution $P(n_1,n_2)$ is not equal to zero only if $n_1+n_2$ is odd. From this fact, $\ket{\boldsymbol{\Phi}_2}_{12}$ is understood as the odd two-mode coherent state \cite{Chai92}. On the other hand, we see that this property is destroyed due to the effect of thermal noise from the case of $\theta=0.5$, in which the dispersion of the photon-number distribution is appeared in the direction of the $n_2$-axis.

\section{Evaluation of the degree of entanglement of $\boldsymbol{\Phi}_2$}
As mentioned in Section 2, the states $\boldsymbol{\Phi}_2$ and $\boldsymbol{\Phi}_4$ are maximally entangled when there is no noise (Eq.(\ref{maximallyentanglement})). Further, these two states are associated with each other by a unitary operator (Eq.(\ref{purecasesymmetryrelations})). 
In this section, we consider the entanglement property of the state $\boldsymbol{\Phi}_2$ with thermal noise.
The analysis method of the entanglement property of $\hat{\boldsymbol{\rho}}_{12}(\boldsymbol{\Phi}_2)$ is based on Ref. 26. 

The fully entangled fraction for the state $\hat{\boldsymbol{\sigma}}_{12}$ is defined as
\begin{equation}
	f(\hat{\boldsymbol{\sigma}}_{12})
	=
	\max
	{}_{12}\bra{\bf e}\hat{\boldsymbol{\sigma}}_{12}\ket{\bf e}_{12},
\end{equation}
where the maximum is taken over all maximally entangled states $\ket{\bf e}_{12}$.
Here we let
\begin{equation}
	f(
	\hat{\boldsymbol{\rho}}_{12}(\boldsymbol{\Phi}_2)
	)
	=
	\max_{\alpha}\Bigl[
	{}_{12}\bra{\boldsymbol{\Phi}_2(\alpha)}
	\hat{\boldsymbol{\rho}}_{12}(\boldsymbol{\Phi}_2)
	\ket{\boldsymbol{\Phi}_2(\alpha)}_{12}
	\Bigr],
	\label{functionf124}
\end{equation}
in order to analyse the state $\hat{\boldsymbol{\rho}}_{12}(\boldsymbol{\Phi}_2)$. 
Note that we chose the state
$
{\ket{\boldsymbol{\Phi}_2(\alpha)}_{12}}
=
{
\mathcal{N}_-(\alpha)
\left[
\ket{\alpha,\alpha}_{12}
-
\ket{-\alpha,-\alpha}_{12}
\right]
}
$
as a maximally entangled state, where $\mathcal{N}_-(\alpha)=1/\sqrt{2(1-\exp[-4|\alpha|^2])}$.
The complete form of Eq.(\ref{functionf124}) is shown in Appendix A. 
By using the resulting value of $f(\hat{\boldsymbol{\rho}}_{12}(\boldsymbol{\Phi}_2))$, a lower bound of the entanglement of formation is given by
\begin{equation}
	E(
	\hat{\boldsymbol{\rho}}_{12}(\boldsymbol{\Phi}_2)
	)
	\geq
	{\sf H}(f(\hat{\boldsymbol{\rho}}_{12}(\boldsymbol{\Phi}_2))),
\end{equation}
where 
\begin{equation}
{\sf H}(f)
=
\left\{
\begin{array}{rl}
h_2(\frac{1}{2}+\sqrt{f(1-f)}), & f\geq 0.5\\
0, & f<0.5
\end{array}
\right.
\end{equation}
with the binary entropy function
$h_2(x)=-x\log_2 x-(1-x)\log_2(1-x)$.

\begin{figure}[t!]
\centerline{
	\includegraphics[scale = 1]{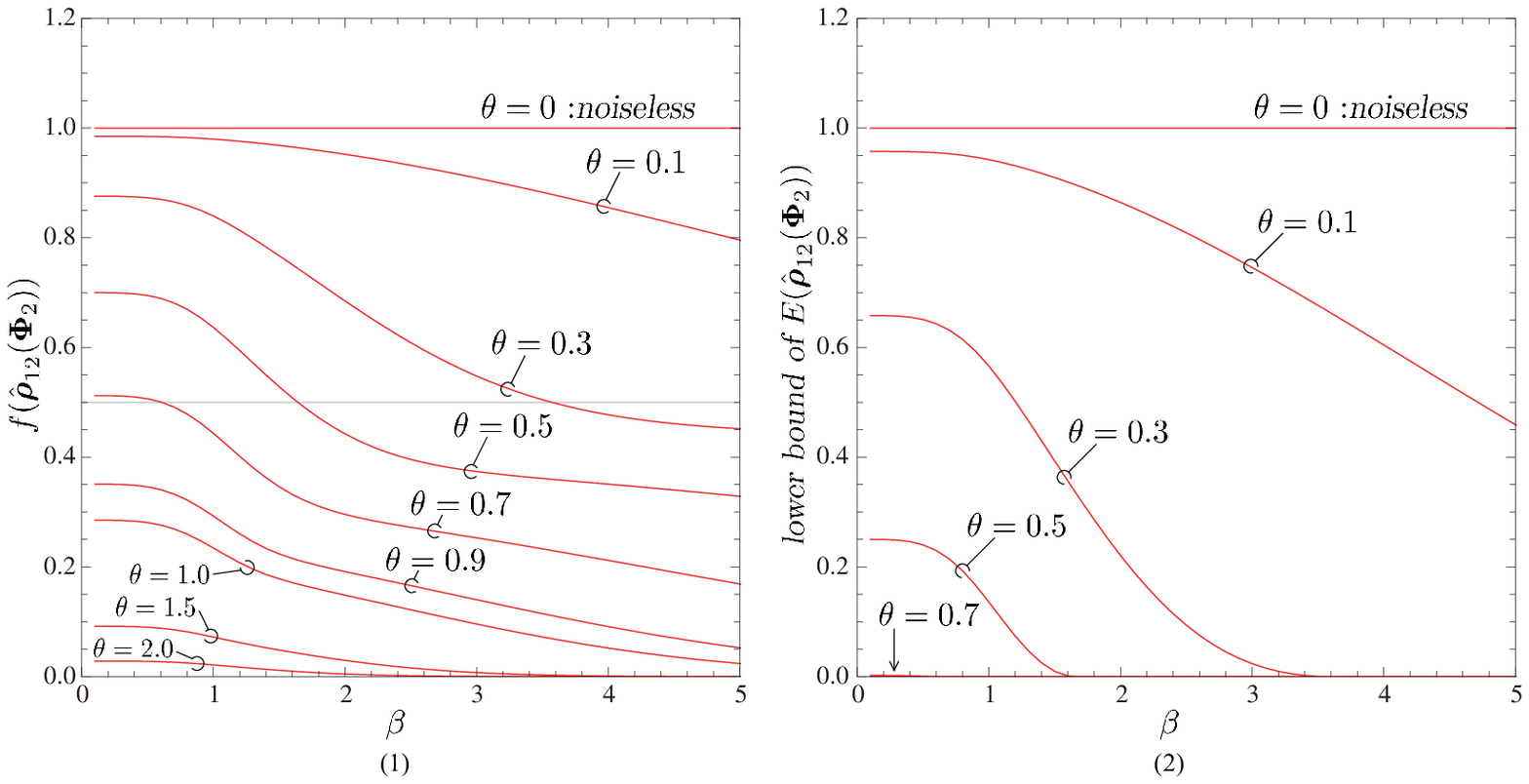}
}
\caption{
Entanglement property. 
(1) behavior of $f(\hat{\boldsymbol{\rho}}_{12}(\boldsymbol{\Phi}_2))$. 
(2) lower bound of $E(\hat{\boldsymbol{\rho}}_{12}(\boldsymbol{\Phi}_2))$ obtained from $f(\hat{\boldsymbol{\rho}}_{12}(\boldsymbol{\Phi}_2))$.
}
\label{fig2}
\vspace{2em}
\hrule
\vspace{2em}
\end{figure}

Figure \ref{fig2} shows the numerical behavior of $f(\hat{\boldsymbol{\rho}}_{12}(\boldsymbol{\Phi}_2))$ and the lower bound of $E(\hat{\boldsymbol{\rho}}_{12}(\boldsymbol{\Phi}_2))$ obtained from the value of $f(\hat{\boldsymbol{\rho}}_{12}(\boldsymbol{\Phi}_2))$. Although it might not be tight, the degradation of the degree of entanglement due to thermal noise is obviously observed through the graph (2) of Figure \ref{fig2}. 
Thus the degree of entanglement is degraded by increasing the amplitude parameter $\beta$ and by increasing the thermal noise parameter $\theta$. The threshold level of $f$ in the function ${\sf H}(f)$ is $0.5$. Hence, from the graph (1) of Figure \ref{fig2}, we roughly estimate that the limit of the thermal noise parameter $\theta$ to keep entanglement is $\sim 0.7$.

\section{Error performance in the minimax discrimination problem}
Suppose there are two quantum states
$\hat{\sigma}_{\sf 0}$ and $\hat{\sigma}_{\sf 1}$, where $\hat{\sigma}_i\geq 0$ and ${\rm Tr}\hat{\sigma}_i=1$. If the probability distribution ${\bf p}=(p_{\sf 0},p_{\sf 1})$ of the states is given in advance, one can employ the quantum Bayes strategy \cite{Hel76,Hol01book}. At that time, the average probability of error of the Bayes optimal discrimination for given ${\bf p}$ is formally written as
\begin{equation}
	\bar{P}_{\rm e}^{\rm min}({\bf p})
	=
	\min_{\mit\Pi}\left[
	p_{\sf 0}
	{\rm Tr}\hat{\mit\Pi}_{\sf 1}\hat{\sigma}_{\sf 0}
	+
	p_{\sf 1}
	{\rm Tr}\hat{\mit\Pi}_{\sf 0}\hat{\sigma}_{\sf 1}
	\right],
\end{equation}
where ${\mit\Pi}=\{\hat{\mit\Pi}_{\sf 0},\hat{\mit\Pi}_{\sf 1}:\hat{\mit\Pi}_{\sf 0}+\hat{\mit\Pi}_{\sf 1}=\hat{1}\ {\rm and}\ \hat{\mit\Pi}_i\geq 0\}$ is a positive operator-valued measure (POVM) which describes a detection profile for the states. 

Clearly $\bar{P}_{\rm e}^{\rm min}({\bf p})$ depends on the choice of $\bf p$. To remove the dependency of choice of the probability distribution $\bf p$, we employ the quantum minimax strategy \cite{HiI82,Kat12a,NKU15}. When the quantum minimax strategy is employed, the average probability of error in the minimax optimal discrimination is given by
\begin{equation}
	\bar{P}_{\rm e}^{\circ}
	=
	\min_{\mit\Pi}\max_{\bf p}
	\left[
	p_{\sf 0}
	{\rm Tr}\hat{\mit\Pi}_{\sf 1}\hat{\sigma}_{\sf 0}
	+
	p_{\sf 1}
	{\rm Tr}\hat{\mit\Pi}_{\sf 0}\hat{\sigma}_{\sf 1}
	\right]
	=
	\max_{\bf p}\min_{\mit\Pi}
	\left[
	p_{\sf 0}
	{\rm Tr}\hat{\mit\Pi}_{\sf 1}\hat{\sigma}_{\sf 0}
	+
	p_{\sf 1}
	{\rm Tr}\hat{\mit\Pi}_{\sf 0}\hat{\sigma}_{\sf 1}
	\right]
	=
	\max_{\bf p}
	\left[\bar{P}_{\rm e}^{\rm min}({\bf p})\right].
\end{equation}
It is not easy to find the optimal probability distribution ${\bf p}^\circ$ in general. However, if the states $\hat{\sigma}_{\sf 0}$ and $\hat{\sigma}_{\sf 1}$ are symmetric, 
the optimal distribution becomes the uniform distribution ${\bf u}=(1/2,1/2)$. 
At that time, we have
$
	\bar{P}_{\rm e}^{\circ}
	=
	\bar{P}_{\rm e}^{\rm min}(1/2,1/2)
$.
Recalling the Helstrom's algorithm for calculating the minimum error probability for binary states \cite{Hel76}, it becomes
\begin{equation}
	\bar{P}_{\rm e}^{\circ}
	=
	\frac{1}{2} - \frac{1}{2}\sum_{\lambda_i\geq 0}\lambda_i
	\label{Helstromalgorithmerrorprobability}
\end{equation}
where $\lambda_i$ are the eigenvalues of 
$\hat{{\sigma}}_{\sf 0}-\hat{{\sigma}}_{\sf 1}$.

\begin{figure}[t!]
\centerline{
	\includegraphics[scale = 1]{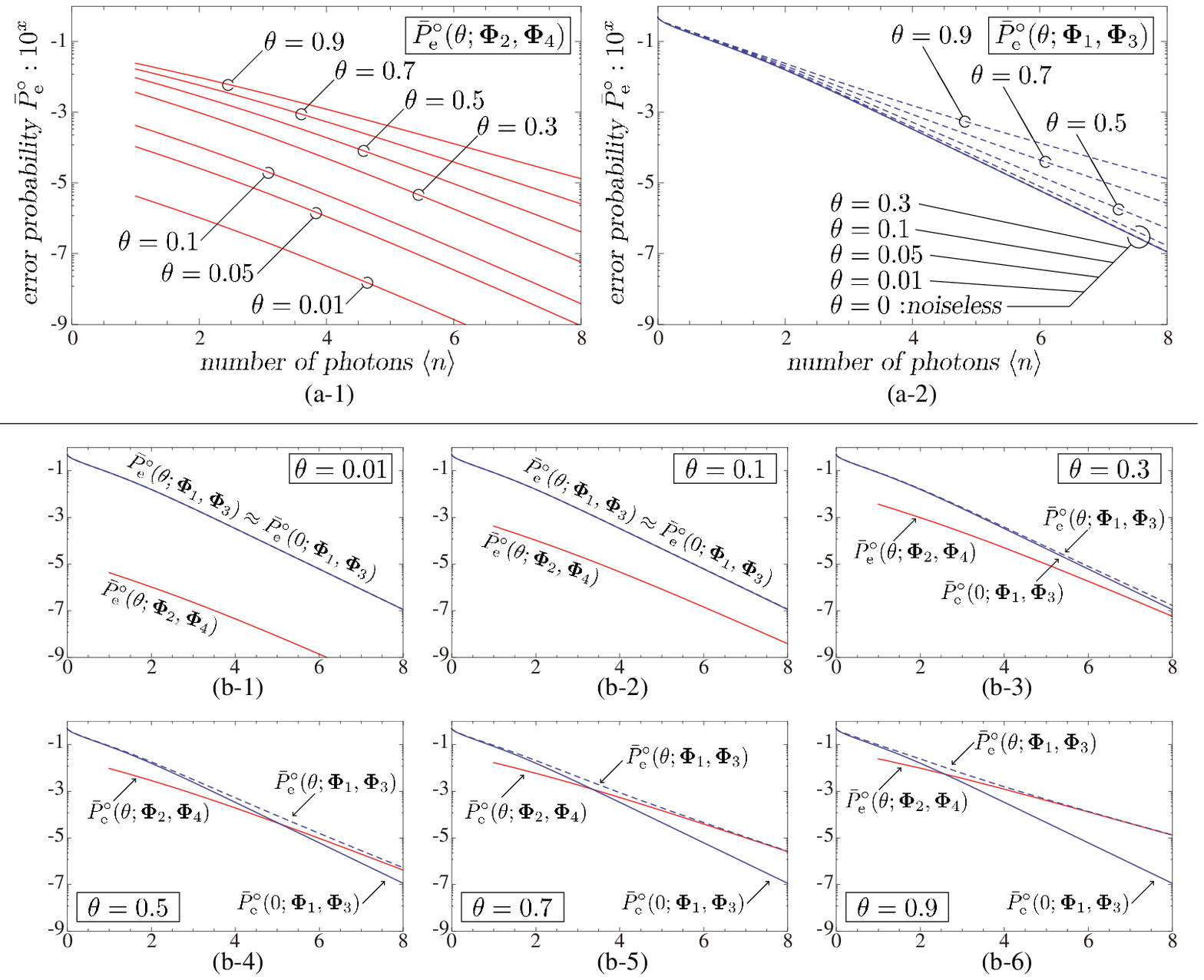}
}
\vspace{1em}
\caption{
Error performance. 
(a-1) $\bar{P}_{\rm e}^\circ(\theta;\boldsymbol{\Phi}_2,\boldsymbol{\Phi}_4)$ for $\theta=0.01,0.05,0.1,0.3,0.5,0.7$, and $0.9$. 
(a-2) $\bar{P}_{\rm e}^\circ(\theta;\boldsymbol{\Phi}_1,\boldsymbol{\Phi}_3)$ for $\theta=0, 0.01,0.05,0.1,0.3,0.5,0.7$, and $0.9$. (b-1) $\bar{P}_{\rm e}^\circ(\theta;\boldsymbol{\Phi}_2,\boldsymbol{\Phi}_4)$ vs $\bar{P}_{\rm e}^\circ(\theta;\boldsymbol{\Phi}_1,\boldsymbol{\Phi}_3)$ at $\theta=0.01$.
(b-2) $\bar{P}_{\rm e}^\circ(\theta;\boldsymbol{\Phi}_2,\boldsymbol{\Phi}_4)$ vs $\bar{P}_{\rm e}^\circ(\theta;\boldsymbol{\Phi}_1,\boldsymbol{\Phi}_3)$ at $\theta=0.1$.
(b-3) $\bar{P}_{\rm e}^\circ(\theta;\boldsymbol{\Phi}_2,\boldsymbol{\Phi}_4)$ vs $\bar{P}_{\rm e}^\circ(\theta;\boldsymbol{\Phi}_1,\boldsymbol{\Phi}_3)$ at $\theta=0.3$.
(b-4) $\bar{P}_{\rm e}^\circ(\theta;\boldsymbol{\Phi}_2,\boldsymbol{\Phi}_4)$ vs $\bar{P}_{\rm e}^\circ(\theta;\boldsymbol{\Phi}_1,\boldsymbol{\Phi}_3)$ at $\theta=0.5$.
(b-5) $\bar{P}_{\rm e}^\circ(\theta;\boldsymbol{\Phi}_2,\boldsymbol{\Phi}_4)$ vs $\bar{P}_{\rm e}^\circ(\theta;\boldsymbol{\Phi}_1,\boldsymbol{\Phi}_3)$ at $\theta=0.7$.
(b-6) $\bar{P}_{\rm e}^\circ(\theta;\boldsymbol{\Phi}_2,\boldsymbol{\Phi}_4)$ vs $\bar{P}_{\rm e}^\circ(\theta;\boldsymbol{\Phi}_1,\boldsymbol{\Phi}_3)$ at $\theta=0.9$.
}
\label{fig3}
\vspace{2em}
\hrule
\vspace{2em}
\end{figure}

Applying the result of Eq.(\ref{Helstromalgorithmerrorprobability}), 
we can compute the error probabilities for the set of the states $\boldsymbol{\Phi}_2$ and $\boldsymbol{\Phi}_4$, and for the set of $\boldsymbol{\Phi}_1$ and $\boldsymbol{\Phi}_3$ in terms of the minimax discrimination problem. Results of the numerical calculation are shown in Figure \ref{fig3}. The graphs (a-1) and (a-2) illustrate $\bar{P}_{\rm e}^\circ(\theta;\boldsymbol{\Phi}_2,\boldsymbol{\Phi}_4)$ and $\bar{P}_{\rm e}^\circ(\theta;\boldsymbol{\Phi}_1,\boldsymbol{\Phi}_3)$, respectively, for $\theta=0.01,0.05,0.1,0.3,0.5,0.7$, and $0.9$. Comparing these two graphs, it is noticed that the behavior of $\bar{P}_{\rm e}^\circ(\theta;\boldsymbol{\Phi}_2,\boldsymbol{\Phi}_4)$ is different from that of $\bar{P}_{\rm e}^\circ(\theta;\boldsymbol{\Phi}_1,\boldsymbol{\Phi}_3)$.
Recall that $\bar{P}_{\rm e}^\circ(\theta;\boldsymbol{\Phi}_2,\boldsymbol{\Phi}_4)=0$ if $\theta=0$. As a value of the thermal noise parameter $\theta$ increases, the error probability $\bar{P}_{\rm e}^\circ(\theta;\boldsymbol{\Phi}_2,\boldsymbol{\Phi}_4)$ rapidly moves upward, while the error probability $\bar{P}_{\rm e}^\circ(\theta;\boldsymbol{\Phi}_1,\boldsymbol{\Phi}_3)$ changes moderately as if to be the case of binary single-mode coherent states $\{\ket{\beta}_1,\ket{-\beta}_1\}$.
The graphs (b-1) to (b-6) involve numerical comparison between $\bar{P}_{\rm e}^\circ(\theta;\boldsymbol{\Phi}_2,\boldsymbol{\Phi}_4)$ and $\bar{P}_{\rm e}^\circ(\theta;\boldsymbol{\Phi}_1,\boldsymbol{\Phi}_3)$ for each $\theta$. In all cases from (b-1) to (b-6), $\bar{P}_{\rm e}^\circ(\theta;\boldsymbol{\Phi}_2,\boldsymbol{\Phi}_4)$ is always smaller than $\bar{P}_{\rm e}^\circ(\theta;\boldsymbol{\Phi}_1,\boldsymbol{\Phi}_3)$. But, the gap between $\bar{P}_{\rm e}^\circ(\theta;\boldsymbol{\Phi}_2,\boldsymbol{\Phi}_4)$ and $\bar{P}_{\rm e}^\circ(\theta;\boldsymbol{\Phi}_1,\boldsymbol{\Phi}_3)$ is getting smaller when the thermal noise parameter $\theta$ increases. When $\theta$ is small enough ($\theta \sim 0.1$ or less), one can easily find a clear gap between $\bar{P}_{\rm e}^\circ(\theta;\boldsymbol{\Phi}_2,\boldsymbol{\Phi}_4)$ and $\bar{P}_{\rm e}^\circ(\theta;\boldsymbol{\Phi}_1,\boldsymbol{\Phi}_3)$ in the region $0\sim \langle n\rangle \sim 8$. When $\theta$ increases, $\bar{P}_{\rm e}^\circ(\theta;\boldsymbol{\Phi}_2,\boldsymbol{\Phi}_4)$ approaches to $\bar{P}_{\rm e}^\circ(\theta;\boldsymbol{\Phi}_1,\boldsymbol{\Phi}_3)$, and finally $\bar{P}_{\rm e}^\circ(\theta;\boldsymbol{\Phi}_2,\boldsymbol{\Phi}_4)$ almost touches to $\bar{P}_{\rm e}^\circ(\theta;\boldsymbol{\Phi}_1,\boldsymbol{\Phi}_3)$. 
\subsection{Conclusions}
The so-called quasi-Bell entangled coherent states, $\boldsymbol{\Phi}_1$, $\boldsymbol{\Phi}_2$, $\boldsymbol{\Phi}_3$ and $\boldsymbol{\Phi}_4$, were studied in the presence of thermal noise. In the analysis, we assumed thermal noise affects only one of the two modes of each state. 
By using the thermalizing operator with thermal noise parameter $\theta$, the matrix representation of the density operators of the quasi-Bell entangled coherent states with thermal noise was derived. 
To seek the entanglement property of the states in the presence of thermal noise, a lower bound of the entanglement of formation for the state $\boldsymbol{\Phi}_2$ was numerically computed. From this, we observed that the range of $\theta$ to keep entanglement of $\boldsymbol{\Phi}_2$ is limited to a small region.
We also considered the optimal quantum discrimination problem for the states $\boldsymbol{\Phi}_2$ and $\boldsymbol{\Phi}_4$ and for $\boldsymbol{\Phi}_1$ and $\boldsymbol{\Phi}_3$. 
From numerical comparison, it is noticed that the behavior of the error probability $\bar{P}_{\rm e}^\circ(\theta;\boldsymbol{\Phi}_2,\boldsymbol{\Phi}_4)$ is different from that of $\bar{P}_{\rm e}^\circ(\theta;\boldsymbol{\Phi}_1,\boldsymbol{\Phi}_3)$, and the gap between $\bar{P}_{\rm e}^\circ(\theta;\boldsymbol{\Phi}_2,\boldsymbol{\Phi}_4)$ and $\bar{P}_{\rm e}^\circ(\theta;\boldsymbol{\Phi}_1,\boldsymbol{\Phi}_3)$ is getting smaller when $\theta$ increases.
The advantage of the use of the states of $\boldsymbol{\Phi}_2$ and $\boldsymbol{\Phi}_4$ against the states of $\boldsymbol{\Phi}_1$ and $\boldsymbol{\Phi}_3$ vanishes when $\theta$ is large. In summary, the effective range for the use of the states of $\boldsymbol{\Phi}_2$ and $\boldsymbol{\Phi}_4$ in information-communication applications would be limited rather than other two, so that careful design for such applications is needed.


\section*{acknowledgment}
The author is grateful to Professor Osamu Hirota of Tamagawa University for his helpful discussions. This work was supported in part by JSPS KAKENHI Grant Number 15K06082.


\appendix
\section{}
Eq.(\ref{functionf124}) is given by 
\begin{eqnarray}
f(\hat{\boldsymbol{\rho}}_{12}(\boldsymbol{\Phi}_2))
=
\max_{\alpha}
\Bigl[\Bigl[
&&
2\frac{\mathcal{N}_-^2(\alpha) \mathcal{N}_-^2(\beta)}{\cosh^2\theta}\times
\Bigl\{
\nonumber\\
&&
+
\exp[-\left|\frac{\alpha}{\cosh\theta}-\beta\right|^2]
\exp[-|\alpha-\beta|^2]
\nonumber\\
&&
-\exp[
-|\alpha|^2(1+\tanh^2\theta)
+\frac{\alpha^\ast\beta}{\cosh\theta}
-\frac{\alpha\beta^\ast}{\cosh\theta}
-|\beta|^2
]
\exp[
-|\alpha|^2
+\alpha^\ast\beta
-\alpha\beta^\ast
-|\beta|^2
]
\nonumber\\
&&
-\exp[
-|\alpha|^2(1+\tanh^2\theta)
-\frac{\alpha^\ast\beta}{\cosh\theta}
+\frac{\alpha\beta^\ast}{\cosh\theta}
-|\beta|^2
]
\exp[
-|\alpha|^2
-\alpha^\ast\beta
+\alpha\beta^\ast
-|\beta|^2
]
\nonumber\\
&&
+
\exp[-\left|\frac{\alpha}{\cosh\theta}+\beta\right|^2]
\exp[-|\alpha+\beta|^2]
\nonumber\\
&&
-
\exp[
-|\alpha|^2\frac{1}{\cosh^2\theta}
+
\frac{\alpha^\ast\beta}{\cosh\theta}
-
\frac{\alpha\beta^\ast}{\cosh\theta}
-
|\beta|^2
]
\exp[
-|\alpha|^2
+\alpha^\ast\beta-\alpha\beta^\ast-|\beta|^2
]
\nonumber\\
&&
+
\exp[
-|\alpha|^2(1+\tanh^2\theta)
+
\frac{\alpha^\ast\beta}{\cosh\theta}
+
\frac{\alpha\beta^\ast}{\cosh\theta}
-
|\beta|^2
]
\exp[
-|\alpha-\beta|^2
]
\nonumber\\
&&
+
\exp[
-|\alpha|^2(1+\tanh^2\theta)
-
\frac{\alpha^\ast\beta}{\cosh\theta}
-
\frac{\alpha\beta^\ast}{\cosh\theta}
-
|\beta|^2
]
\exp[
-|\alpha+\beta|^2
]
\nonumber\\
&&
-
\exp[
-|\alpha|^2\frac{1}{\cosh^2\theta}
-
\frac{\alpha^\ast\beta}{\cosh\theta}
+
\frac{\alpha\beta^\ast}{\cosh\theta}
-
|\beta|^2
]
\exp[
-|\alpha|^2
-\alpha^\ast\beta+\alpha\beta^\ast-|\beta|^2
]
\Bigr\}
\quad
\Bigr]\Bigr].
\nonumber
\\
\end{eqnarray}
This maximization problem with respect to $\alpha$ was numerically solved to draw the graph (1) of Figure \ref{fig2} under the restriction that all parameter is real.

\begin{thebibliography}{99}
\bibitem{San12}
Sanders, B.C.,
``Review of entangled coherent states,"
J. Phys. A: Math. Theor. 45(24), 244002 (2012).
\bibitem{San13}
Sanders, B.C,
``Forty-five years of entangled coherent states,"
Proceedings of the first international workshop on entangled coherent states and its application to quantum information science (Edited by Usuda, T.S. and Kato, K.), 111-113 (2013). 
\bibitem{HiS01}
	Hirota, O. and Sasaki, M.,
	``Entangled state based on nonorthogonal state,"
	Proc. QCM\&C-Y2K (Edited by Tombeshi, P. and Hirota, O.), 359-366 (2001).
\bibitem{EnH01}
	van Enk, S.J. and Hirota, O.,
	``Entangled coherent states: Teleportation and decoherence,"
	Phys. Rev. A 64(2), 022313 (2001).
\bibitem{Wan01}
Wang, X., 
``Quantum teleportation of entangled coherent states,"
Phys. Rev. A 64(2), 022302 (2001).

\bibitem{JKL01}
Jeong, H., Kim, M.S. and Lee, J., 
``Quantum-information processing for a coherent superposition state via a mixed entangled coherent channel,"
Phys. Rev. A 64(5), 052308 (2001).

\bibitem{JeK02}
Jeong, H. and Kim, M.S.,
``Efficient quantum computation using coherent states,"
Phys. Rev. A 65(4), 042305 (2002).

\bibitem{Hir11}
	Hirota, O., 
	``Error free quantum reading by quasi Bell state of entangled coherent states,"
	e-print arXiv:quant-ph/1108.4163v2 (2011).
\bibitem{KaH13} 
	Kato, K. and Hirota,O., 
	``Effect of decoherence in quantum reading with phase shift keying signal of entangled coherent states,"
	Proc. SPIE 8875, 88750P(2013).


\bibitem{HiI82}
	Hirota, O. and Ikehara, S.,
	``Minimax strategy in the quantum detection theory and its application to optical communications," Trans. IECE of Japan E65, 627-633 (1982).
\bibitem{Kat12a}
	Kato, K., ``Necessary and sufficient conditions for minimax strategy in quantum signal detection," 2012 IEEE Int. Sympo. Inform. Theory (ISIT) Proc., 1082-1086 (2012).
\bibitem{NKU15} 
	Nakahira, K., Kato, K. and Usuda, T.S., 
	``Generalized quantum state discrimination problems,"
	Phys. Rev. A 91(5), 052304 (2015).
\bibitem{BaP89}
	Barnett, S.M. and Phoenix, S.J.,
	``Entropy as a measure of quantum optical correlation,"
	Phys. Rev. A 40, 2404-2409 (1989).
\bibitem{BBPS96}
	Bennett, C.H., Bernstein, H.J., Popescu, S. and Schumacher, B., 
	``Concentrating partial entanglement by local operations,"
	Phys. Rev. A 53, 2046-2052 (1996).
\bibitem{TaU75}
	Takahashi, Y. and Umezawa, H., 
	``Thermo field dynamics,"
	Collect. Phenom. 2, 55-80 (1975).
\bibitem{BaK85}
	Barnett, S.M. and Knight, P.L., 
	``Thermofield analysis of squeezing and statistical mixtures in quantum optics," J. Opt. Soc. Am. B, 2(3), 467-479 (1985).
\bibitem{FeC88}
	Fearn, H. and Collett M.J.,
	``Representation of squeezed states with thermal noise,"
	J. Mod. Opt. 35(3), 553-564 (1988).
\bibitem{MaR89}
	Mann, A. and Revzen, M.,
	``Thermal coherent states,"
	Phys. Lett. A 134(5), 273-275 (1989).
\bibitem{OVMR91}
	Oz-Vogt, J., Mann, A. and Revzen, M.,
	``Thermal coherent states and thermal squeezed states,"
	J. Mod. Opt. 38(12), 2339-2347 (1991).
\bibitem{KMRU89}
	Kireev, A., Mann, A., Revzen, M. and Umezawa, H.,
	``Thermal squeezed states in thermo field dynamics and quantum and thermal fluctuations," Phys. Lett. A 142(4/5), 215-221 (1989).
\bibitem{Lee90}
Lee, C.T., 
``Two-mode squeezed states with thermal noise,"
Phys. Rev. A 42(7), 4193-4198 (1990).
\bibitem{Lu99}
	Lu, W.F., 
	``Thermalized displaced and squeezed number states in the coordinate representation," J. Phys. A: Math. Gen. 32(27), 5037 (1999).
\bibitem{Lu00}
	Lu, W.F.,
	``Thermalized displaced squeezed thermal states,"
	J. Phys. A: Math. Gen. 33(3), 479 (2000).
\bibitem{SKS94}
	Selvadoray, M., Kumar, M.S. and Simon, R.,
	``Photon distribution in two-mode squeezed coherent states with complex displacement and squeeze parameters,"
	Phys. Rev. A, 49(6), 4957-4967 (1994).
\bibitem{Chai92}
	Chai, C.L., 
	``Two-mode nonclassical state via superpositions of two-mode coherent states,"
	Phys. Rev. A 46(11), 7187-7191 (1992).
\bibitem{HENSK01}
	Hirota, O., van Enk, S.J., Nakamura, K., Sohma, M. and Kato, K.,
	``Entangled nonorthogonal states and their decoherence properties,"
	e-print arXiv:quant-ph/0101096v1 (2001).
\bibitem{Hel76}
	Helstrom, C. W.,
	Quantum Detection and Estimation Theory, Academic Press, New York, (1976).
\bibitem{Hol01book}
	Holevo, A.S.,
	Statistical Structure of Quantum Theory, Springer, Berlin, (2001).
\end{thebibliography}
\end{document}